\documentclass[aps,prl,reprint,groupedaddress,showpacs]{revtex4-1}
\usepackage{graphicx}
\usepackage{amsmath}
\usepackage{amssymb}
\usepackage{hyperref}

\begin{document}

\title{Generalized unitary Bogoliubov transformation that breaks fermion number parity}
\author{Jonathan E. Moussa}
\email[]{godotalgorithm@gmail.com}
\affiliation{Sandia National Laboratories, Albuquerque, NM 87185, USA}
\date{\today}

\begin{abstract}
The standard Bogoliubov transformation is generalized to enable fermion number parity breaking.
The new transformation can diagonalize fermion Hamiltonians that
 are quadratic in fermion and number parity operators.
This new variational freedom cannot lower the energy of Hamiltonians that conserve number parity.
However, it has value as a minimal mechanism for continuously varying between states of different fermion number,
 regardless of number parity.

\end{abstract}


\maketitle

The Bogoliubov transformation is a well-established mechanism for generating wavefunctions
 of indefinite fermion number.
However, these wavefunctions continue to have definite number parity.
A generalization that can break number parity exists \cite{generalB},
 but it loses the simple structure of a linear transformation between elementary fermion operators.
Preserving this structure is desirable for theoretical considerations of number parity violations
 on equal footing with number conservation violations.

A new generalization of the Bogoliubov transformation is defined by a set of quadratic fermion Hamitonians,
\begin{align}\label{H}
 \hat{H} &= h_{ij} \hat{c}_i^\dag \hat{c}_j +  \tfrac{1}{2}g_{ij} \hat{c}_i^\dag \hat{c}_j^\dag
 + \tfrac{1}{2}g_{ji}^* \hat{c}_i \hat{c}_j + \tfrac{1}{2}f_i \hat{c}_i^\dag \hat{\pi} + \tfrac{1}{2}f_i^* \hat{\pi} \hat{c}_i \notag \\
 \hat{\pi} &= \exp(i \pi \hat{n}_j), \ \ (\hat{\pi} \hat{c}_I = - \hat{c}_I \hat{\pi} \ \ \& \ \ \hat{\pi}^2 = 1 ),
\end{align}
 in second quantization notation
 with $\hat{n}_I \equiv \hat{c}_I^\dag \hat{c}_I$, $h_{IJ} = h_{JI}^*$, and $g_{IJ} = -g_{JI}$.
Lowercase indices are summed implicitly over all fermion sites.
This model includes a fermion number parity operator, $\hat{\pi}$, with the algebraic properties of a Majorana fermion operator
 but without a corresponding site within the Hilbert space.

As in the Bogoliubov case with $f_I = 0$, calculations are enabled by
 an operator set that forms a Lie algebra,
\begin{equation}
\mathcal{Q} = \{1,\hat{\pi} \hat{c}_I,\hat{c}_I^\dag \hat{\pi},\hat{c}_I \hat{c}_J , \hat{c}_I^\dag \hat{c}_J , \hat{c}_I^\dag \hat{c}_J^\dag \}.
\end{equation}
Any Hermitian operator spanned by this set, $\hat{H} = \hat{X} + \hat{X}^\dag$ with $\hat{X} \in \mathrm{span}(\mathcal{Q})$,
 can be diagonalized by an operator of the form: $\hat{U} = \exp(\hat{Y} - \hat{Y}^\dag)$ with $\hat{Y} \in \mathrm{span}(\mathcal{Q})$.
The transformation applied to the basic fermion operators is
\begin{equation}\label{trans}
 \left[ \begin{array}{c} \hat{U}^\dag \hat{c}_I^\dag \hat{U} \\ \hat{U}^\dag \hat{c}_I \hat{U} \\ \hat{U}^\dag \hat{\pi} \hat{U} \end{array}\right]
 = \underbrace{\left[ \begin{array}{ccc} \xi_{Ij}^* & \phi_{Ij}^* & \gamma_I^* \\ \phi_{Ij} & \xi_{Ij} & \gamma_I \\ \beta_j^* & \beta_j & \alpha \end{array} \right]}_{\mathbf{Z}}
 \left[ \begin{array}{c} \hat{c}_j^\dag \\ \hat{c}_j \\ \hat{\pi} \end{array}\right].
\end{equation}
$\mathbf{Z}$ is constrained to have SO$(2n+1)$ group structure,
\begin{equation}
 \mathbf{Z} = \mathbf{S}^{-1} \mathbf{Q} \mathbf{S} , \ \ \mathbf{S} = \frac{1}{\sqrt{2}}\left[ \begin{array}{ccc} 1 & 1 & 0 \\  -i & i & 0 \\ 0 & 0 & 1 \end{array} \right] , \ \ \mathrm{det}(\mathbf{Q}) = 1,
\end{equation}
 for a real orthogonal matrix $\mathbf{Q}$.
If $\mathbf{Q}$ solves $\mathbf{H} \mathbf{Q} = \mathbf{Q} \mathbf{E}$,
 which is a real antisymmetric eigenvalue problem \cite{as_form} with
\begin{align}
\mathbf{H} &= \left[\begin{array}{ccc} \mathrm{Im}(g_{IJ}+h_{IJ}) & \mathrm{Re}(g_{IJ}-h_{IJ}) & \mathrm{Im}(f_I) \\
\mathrm{Re}(h_{IJ}+g_{IJ}) & \mathrm{Im}(h_{IJ}-g_{IJ}) & \mathrm{Re}(f_I) \\
  -\mathrm{Im}(f_J) & -\mathrm{Re}(f_J) & 0 \end{array} \right]  \notag \\
\mathbf{E} &= \left[\begin{array}{ccc} 0 & -\epsilon_I \delta_{IJ} & 0 \\ \epsilon_I \delta_{IJ} & 0 & 0 \\ 0 & 0 & 0 \end{array}\right] , \ \ \epsilon_I \ge 0,
\end{align}
then $\hat{U}$ diagonalizes $\hat{H}$: $\hat{U}^\dag \hat{H} \hat{U} = \tfrac{1}{2} ( h_{ii} - \epsilon_i) + \epsilon_i \hat{n}_i$.
The transformed vacuum state, $\hat{U}| \varnothing \rangle$, is the ground state.

The result can be factored into $\hat{U} = \hat{U}_\mathrm{B} \hat{U}_\theta$,
 where $\hat{U}_\mathrm{B}$ is a standard Bogoliubov transformation ($\beta_I = \gamma_I = 0$) and
 $\hat{U}_\theta = \exp(\theta \hat{\pi} \hat{c} - \theta \hat{c}^\dag \hat{\pi})$
 breaks number parity by coupling to a single fermion mode, $\hat{c} := \beta_i \hat{c}_i$, which acts as
\begin{subequations} \begin{align}
 \hat{U}_\theta^\dag \hat{c} \hat{U}_\theta &=  - \hat{c}^\dag \sin^2 \theta + \hat{c} \cos^2 \theta - \tfrac{1}{2} \hat{\pi} \sin 2 \theta \\
 \hat{U}_\theta^\dag \hat{\pi} \hat{U}_\theta &=  (\hat{c}^\dag + \hat{c}) \sin 2 \theta +  \hat{\pi} \cos 2 \theta .
\end{align} \end{subequations}
The transformed vacuum that results is a superposition of even and odd number parity states,
\begin{equation}\label{wave}
 \hat{U} | \varnothing \rangle = \hat{U}_\mathrm{B} | \varnothing \rangle \cos \theta
 - \hat{U}_\mathrm{B} \hat{c}^\dag | \varnothing \rangle  \sin \theta .
\end{equation}
For a Hamiltonian that conserves number parity, there is no matrix element between the even and odd states,
 which means that the variational ground state will have definite number parity.
Multiconfigurational calculations where each wavefunction has freedom to independently break number parity \cite{multibreak}
 will also have this null outcome.
There remains the possibility that nonvariational models of quantum correlation
 may artificially stabilize a parity-broken ground state.

Number parity breaking can be a useful theoretical tool
 even when a parity-broken ground state is not stable.
For example, consider a system where the Hartree-Fock and exact ground states
 have different number parity,
\begin{equation}
 \hat{H}_4 =  | \delta | \hat{n}_1 + \hat{\Delta}
  + V (\hat{c}_4^\dag \hat{c}_3^\dag \hat{c}_2 \hat{c}_1+ \hat{c}_1^\dag \hat{c}_2^\dag \hat{c}_3 \hat{c}_4),
\end{equation}
 for $\hat{\Delta} = \tfrac{1}{3} | \Delta | (1 - \hat{n}_2 + \hat{n}_3 + \hat{n}_4)$.
If $\hat{U}_\theta$ with $\hat{c} = \hat{c}_1$ is applied to
 the Hartree-Fock ground state, $\hat{c}_2^\dag | \varnothing \rangle$,
 its total energy increases to a value of
\begin{equation}
 E_\mathrm{HF}(\theta) = | \delta | \sin^2 \theta.
\end{equation}
For a post-Hartree-Fock state of the form $\hat{U}_2 \hat{U}_\theta \hat{c}_2^\dag | \varnothing \rangle$
 with a unitary transformation, $\hat{U}_2$, that doesn't mix number sectors,
 the variational ground state energy is lowered to
\begin{equation}
 E_\mathrm{var}(\theta) =  \left[ \tfrac{| \delta | + | \Delta | }{2} - \sqrt{\left(\tfrac{| \delta | + | \Delta | }{2}\right)^2 + V^2} \right] \sin^2 \theta.
\end{equation}
The $\theta=0$ correlated ground state is unstable to parity breaking
 even though the minimum energy occurs for a state of fixed number.
This avoids the need to consider multiple number-constrained Hartree-Fock solutions
 as reference states for correlation models based on a pure state.
For correlation models based on a mixed state, orbital occupations
 can be continuously varied without breaking number parity.

In conclusion, the Bogoliubov transformation can be generalized
 to unify the breaking of number conservation and number parity.
The immediate consequences of this generalization are modest,
 but it may have applications to improving the stability of quantum correlation models based on a pure reference state.

\begin{acknowledgments}
Sandia National Laboratories is a multi-program laboratory managed and operated by Sandia Corporation,
a wholly owned subsidiary of Lockheed Martin Corporation, for the U.S. Department of Energy's
National Nuclear Security Administration under contract DE-AC04-94AL85000.

Thanks to Jay Deep Sau, Toby Jacobson, and Rick Muller for helpful discussions.
\end{acknowledgments}

\end{document}